\documentclass[]{interact}

\usepackage{epstopdf}
\usepackage[caption=false]{subfig}

\usepackage[numbers,sort&compress]{natbib}
\bibpunct[, ]{[}{]}{,}{n}{,}{,}
\makeatletter
\def\NAT@def@citea{\def\@citea{\NAT@separator}}
\makeatother

\theoremstyle{plain}

\theoremstyle{definition}

\theoremstyle{remark}

\begin{document}

\articletype{FULL ARTICLE}

\title{Structural, Elastic, Electronic and Optical Properties of $Be_2X(X=C, Si, Ge, Sn)$: First Principle Study}

\author{
\name{Prakash Chanda Gupta\textsuperscript{a}\thanks{Corresponding Author Prakash Chanda Gupta. Email: pcgupta.physics@gmail.com} and Rajendra Adhikari\textsuperscript{b}}
\affil{\textsuperscript{a} \textsuperscript{b}Department of Physics, Kathmandu University, Nepal}
}

\maketitle

\begin{abstract}
We computed structural, elastic, electronic and optical properties of $Be_2X(X=C, Si, Ge, Sn)$ family of antifluorite with ab initio DFT calculations using the generalized gradient approximation (GGA). The different parameters such as geometry optimization, band structure, density of states, elastic constants, dielectric functions have been studied. We also calculated bandgap using PBE0 and HSE hybrid functionals to compare experimental bandgap of $Be_2C$. Although three of the compounds are hypothetical in nature, their formation energy found to be negative. The calculated values of elastic constants indicates antifluorite $Be_2X$ are mechanically stable. The graph of real part of epsilon shows negative value giving promising result for blanket behaviour of $Be_2X$ from radiation damage.
\end{abstract}


\section{Introduction}

Antifluorite $Be_2X(X=C,Si,Ge,Sn)$ structure form the simplest family of metal - semiconductor (Group II-IV elements) hybrid materials. Antifluorite structure has a similar structure to diamond. Group IV atoms occupy fcc sites with eightfold coordination and II atoms are at tetrahedral sites with fourfold coordination of resulting structure of cubic Fm-3m space group.\\
The first member of the family Beryllium carbide is colourless in pure state and is a transparent crystalline solid and very hard compound, like diamond, a pure carbon compound \cite{C39890001652}. $Be_2C$ is a refracting material. In presence of moisture, it is chemically instable. It is hard and have large elastic constants. The sound velocity in $Be_2C$ is high \cite{C9RA01573F}. It has a high melting point and a good thermal conductivity \cite{1954}. It is observed that $Be_2C$ is a bit harder than SiC \cite{1952}. $Be_2C$ is also resistance to radiation damage and may be used in fission reactor components and as a blanket material in fusion reactor \cite{KLEYKAMP200188}. \\
The $Be_2C$ is among the few known alkaline earth methanides. The $Be_2C$ used in ceramic and nuclear technology, has attracted quite a few theoretical as well as experimental studies \cite{C9RA01573F}. $Be_2C$ is highly poisonous and therefore only few experimental research has been done. The three compounds $Be_2X(X=Si, Ge, Sn)$ are hypothetical materials and are not known to exit in bulk at present. But some experiments strongly suggests existence of clusters $(Be_2Si_n)$ on the surface of Silicon \cite{771086,article}. So, there are very few study on these materials yet. The goal of this paper is to obtain the basic structure, electronic and optical properties of $Be_2X$ by DFT \cite{sholl2011density}.
\section{Computational Methods}\label{class}
Calculations were carried out through DFT package Quantum Espresso (version 6.7) using pseudopotential plane-wave method \cite{singh2006planewaves}. The PBE-PAW pseudopotentials were used with an energy cutoff of 50 $Ry$ and k-point mesh 6 x 6 x 6 is employed and further increased to 14 x 14 x 14 in calculating electronic and optical properties. The lattice vectors were optimized with total energy converging within $10^{-4}$ a.u. Non-conserving pseudopotentials are used for calculations of optical properties and Projector Augmented Waves pseudopotentials are used for all other calculations. PBE exchange-correlation functional of GGA \cite{PhysRevLett.77.3865} is used in all calculations. In the calculations, the $Be (2s^2), C (2s^2 2p^2), Si (3s^2 3p^2), Ge (4s^2 4p^2)$ and $Sn (5s^2 5p^2)$ states are treated as valence electrons. Hybrid functionals PBE0 \cite{doi:10.1063/1.472933} and HSE \cite{doi:10.1063/1.1564060} are also used for calculating Bandgap of the four structure to incorporate a portion of exact exchange from Hatree-Fock theory  with rest of exchange-correlation energy from ab initio calculations. \\ 
The elastic constants were calculated out through ElaStic code \cite{GOLESORKHTABAR20131861} using Energy-strain method and the maximum strain value was 0.03$\%$.The optical properties were determined using many-body effects solving Bethe-Salpeter Equations for accurate theoretical description and compared with Independent Particle Approximation \cite{PhysRevB.48.11705} using YAMBO code \cite{2019} from the complex dielectric function, $\varepsilon (\omega) = \varepsilon_1 (\omega) + i \varepsilon_2 (\omega)$. \\
The crystal structure is shown in figure 1. It has cubic space group number 225 where Be atom occupy $8c$ and X atom occupy $4a$ Wyckoff positions \cite{RUSCHEWITZ2003115}.
\begin{figure}[!h]
	\centering
	\includegraphics[width=0.6\linewidth]{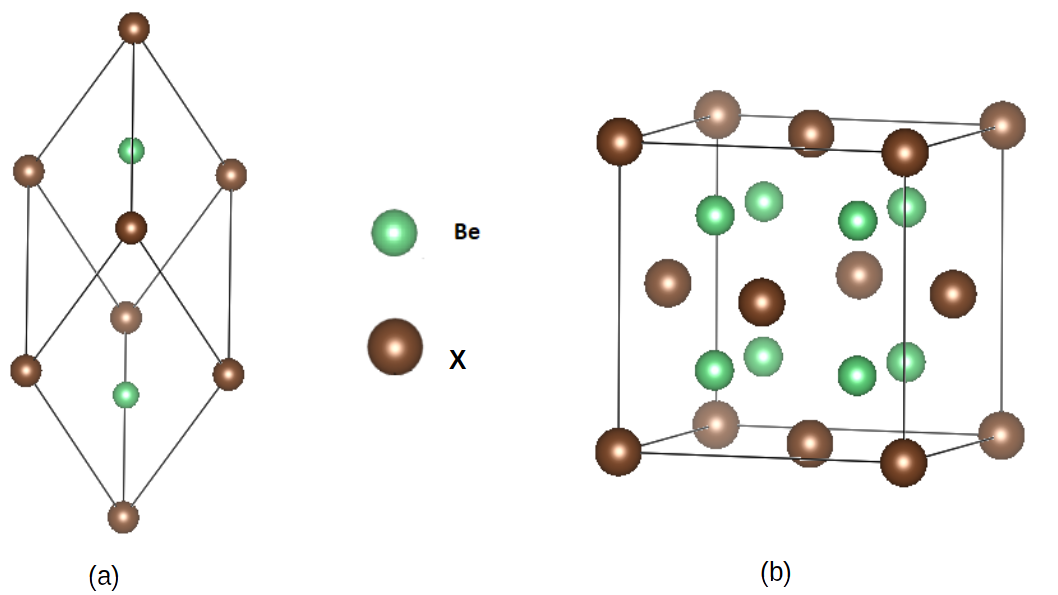}
	\caption{Primitive cell (a) and Cubic crystal structure (b) model of $Be_2X$}
\end{figure}
\begin{table}[h]
	\centering
	\caption{Atomic Positions of $Be_2X$}
	\begin{tabular}{p{2cm}p{2cm}p{6cm}}
		\toprule
		Site & Location & Co-ordinates\\
		\midrule
		Be & 8c & (0.25, 0.25, 0.25),(0.75, 0.75, 0.75) \\
		\\
		X & 4a & (0, 0, 0) \\
		\bottomrule
	\end{tabular}
\end{table}
\section{Results}
\subsection*{Structural Properties}
The antifluorite has cubic structure. The space group of antifluorite structure is Fm-3m with space group number 225. The structure consists of 4 molecules per unit cell. The atomic positions are given in Table 1. In order to analyze structural properties of $Be_2X$, cut-off energy and k-points optimizations were performed followed by geometry optimization. The Total energy vs. Volume (Appendix A) graphs have been plotted coupled with Murnaghan equation of state \cite{Murnaghan1944} for the four structures. The minimum total energy corresponds to the stable point of the crystal. The lattice constant parameter at this point and the Bulk modulus for the four structures is listed in Table 2 and Table 3 with available previous similar work and experimental data. \\
The calculated lattice constant for $Be_2C$ using GGA is in good agreement with the experimental data. The experimental data are not available for other three structures. The calculated Bulk modulus value is also found to be in good agreement with other GGA calculations \cite{doi:10.1063/1.4810267, Yan2011} but lower than the LDA results \cite{PhysRevB.51.10392,PhysRevB.48.17138}. This may be due to over binding tendency of Local density approximation (LDA). \cite{PhysRev.136.B864}. 
\begin{table}[h]
	\centering
	\caption{Lattice Constant a of $Be_2X(X=C,Si,Ge,Sn)$}
	\begin{tabular}{p{4cm}p{1.8cm}p{1.5cm}p{1.5cm}p{1.5cm}p{1.5cm}p{1.5cm}}
		\toprule
		Work & Method  & $Be_2C$ & $Be_2Si$ & $Be_2Ge$ & $Be_2Sn$ \\
		\midrule
		This work & GGA & 4.327 \AA & 5.268 \AA & 5.366 \AA & 5.799 \AA \\
		Paliwal et al. \cite{doi:10.1063/1.4810267}  & GGA & 4.335 \AA  \\
		Yan et al. \cite{Yan2011}  & GGA & 4.33 \AA & 5.28 \AA  \\
		Lee et al. \cite{PhysRevB.51.10392} & LDA & 4.27 \AA & 5.22 \AA \\
		Corkill and Cohen \cite{PhysRevB.48.17138} & LDA-CA & 4.23 \AA & 5.18 \AA \\
		Experimental \cite{RUSCHEWITZ2003115} & -- & 4.330 \AA   \\
		\bottomrule
	\end{tabular}
\end{table}
\begin{table}[h]
	\centering
	\caption{Bulk modulus $B_o$ of $Be_2X(X=C,Si,Ge,Sn)$}
	\begin{tabular}{p{4cm}p{1.8cm}p{1.8cm}p{1.8cm}p{1.8cm}p{1.8cm}}
		\toprule
		Work & Method  & $Be_2C$ & $Be_2Si$ & $Be_2Ge$ & $Be_2Sn$ \\
		\midrule
		This work & GGA & 198.7 GPa & 97.4 GPa & 91.5 GPa & 72.1 GPa \\
		Paliwal et al. \cite{doi:10.1063/1.4810267}  & GGA  & 198.9 GPa \\
		Yan et al. \cite{Yan2011}  & GGA & 195.8 GPa & 94.5 GPa \\
		Lee et al. \cite{PhysRevB.51.10392} & LDA & 216.0 GPa & 102.5 GPa \\
		Corkill and Cohen \cite{PhysRevB.48.17138} & LDA-CA & 216 GPa & 103.2 GPa \\
		Experimental \cite{RUSCHEWITZ2003115} & --   \\
		\bottomrule
	\end{tabular}
\end{table}
\subsection*{Elastic Properties}
The stiffness matrices [$C_{ij}$] is calculated using PBE functionals and Energy-strain method. There are three independent elastic constants $C_{11}, C_{12}$ and $C_{44}$ for a cubic system. These elastic constants are used to calculate elastic property terms through the Viogt-Reuss-Hill (VRH) averaging scheme for cubic system \cite{Man2011, Varshney2015}. The different scheme shear modulus G as well as the bulk modulus B are related to [$C_{ij}$] as,  \\
\begin{align*}
	G_v &= \frac{C_{11} - C_{12} + 3 C_{44}}{5} \\ \\
	G_R & = \frac{5C_{44}(C_{11} - C_{12})}{4C_{44} + 3(C_{11} - C_{12})}  \\ \\
	G &= \frac{1}{2} (G_v + G_R) \\ \\
	B_v &= B_R = \frac{C_{11} + 2C_{12}}{3} \\ \\ 
	B &= \frac{1}{2} (B_v + B_R)
\end{align*}
Young's modulus E and Poisson's ratio $v$ are also calculated from the relations,\\
\begin{align*}
	E &= \frac{9BG}{(3B+G)} \\ \\
	v &= \frac{3B - 2G}{2(3B+G)}
\end{align*}
Their calculated value is shown in Table 4. The calculated value of elastic constants satisfy the Born's stability criteria for the cubic system \cite{PhysRevB.90.224104}, i.e.,$C_{11}-C_{12} > 0, C_{11}+2C_{12} > 0$ and $C_{44}>0$ for all four structures. This shows that $Be_2X$ is mechanically stable. The bulk modulus $B_o$ calculated from fitting the Murnaghan equation of states and the bulk modulus B obtained
from the above Voigt–Reuss–Hill approximation have comparable value. This shows that our elastic calculations are consistent and reliable. \\
For $Be_2C$, the value of shear modulus is large. This indicates that it can withstand shear strain to a large extent. This assures the use of $Be_2C$ as a hard material in technological applications. The ratio of Bulk modulus to the Shear modulus is very low which indicated fragile nature of $Be_2C$. The Poisson's ratio is calculated very less and shows high brittle nature of $Be_2C$. The calculated elastic constants for $Be_2X(X=Si, Ge, Sn)$ are much smaller than $Be_2C$. This may be due to metallicity of these compounds.
\begin{table}[h!]
	\centering
	\caption{Calculated Values of Elastic Constants ($C_{ij}$), bulk modulus (B), shear modulus (G), Young's modulus (E) and Poisson's Ratio ($v$) of $Be_2X(X=C, Si, Ge, Sn)$}
	\begin{tabular}{p{1.5cm}p{1.5cm}p{1.5cm}p{1.5cm}p{1.5cm}p{1.5cm}p{1.5cm}}
		\toprule
		Parameter  & $Be_2C$  [GPa] & $Be_2C ^a$ [GPa] & $Be_2Si$ [GPa] & $Be_2Si ^a $ [GPa] & $Be_2Ge$ [GPa] & $Be_2Sn$ [GPa]  \\
		\midrule
		$C_{11}$ & 579.6 & 570.5  & 166.9  & 133.0 & 148.6  & 103.4 \\
		$C_{12}$ & 13.3  & 16.0 & 73.5  & 82.5 & 69.2 & 59.8 \\
		$C_{44}$ & 200.9  & 208.7 & 106.0 & 99 & 95.4 & 49.9 \\
		B & 202.05  & 200.8 & 104.63 & 99.3 & 95.65  & 74.37 \\
		G & 230.56  & 233.8 & 76.29  & 57.6 & 67.09  & 35.77 \\
		E & 501.08  & 505.4 & 184.11 & 154.8 & 163.13  & 92.49 \\
		$v$ & 0.09 & 0.081 & 0.21 & 0.26 & 0.22 & 0.29 \\
		\bottomrule
		\vspace{0.4cm}
	\end{tabular}
	\raggedright
	$^a$ Using DFT with GGA.[ Yan et al.]\cite{Yan2011}
\end{table}
\newpage
\subsection*{Electronic Properties}
The band structure, total and projected density of states (DOS) \cite{harrison2012electronic} were obtained using DFT with generalized gradient approximation. The energy band structure  along principle symmetry points obtained is shown in Figure 2 with highest valence band set to origin. We demonstrated 6 conducting bands for $Be_2C$ \& $Be_2Si$ and  5 conducting bands for $Be_2Ge$ \& $Be_2Sn$ . The valence band can be classified into two group. The lower energy band is mainly due to s states of Group IV atoms and p states of Be atom . The remaining group of bands is mainly due to p states of group IV atoms along with the p states of Be atom. The conduction bands includes s and p states. This is shown in partial DOS Figure 3. \\ \\
For $Be_2C$, the indirect band gap is obtained to be 1.170 eV and direct band gap is 4.132 eV. The result is close to experimental data \cite{PhysRevB.58.6837}. This classify $Be_2C$ as a semiconductor with a mid band gap. The calculated Total DOS shows the contribution of valence electrons from below 10 eV and conduction electron contributing to DOS above 10 eV. However, the other three structures have finite density of states at the Fermi level. This indicates they have metallic behaviour in its crystalline state. Thus, these structures are significantly different from $Be_2C$ at the Fermi Energy level though they are assumed to have same crystal structure. This discrepancy may be due to overlaping of other IV atom outer $p$ orbitals with neighbour Be atoms than the C atom $2p$ orbitals. \\
\newpage
\begin{figure}[h!]
	\includegraphics[width=1\linewidth]{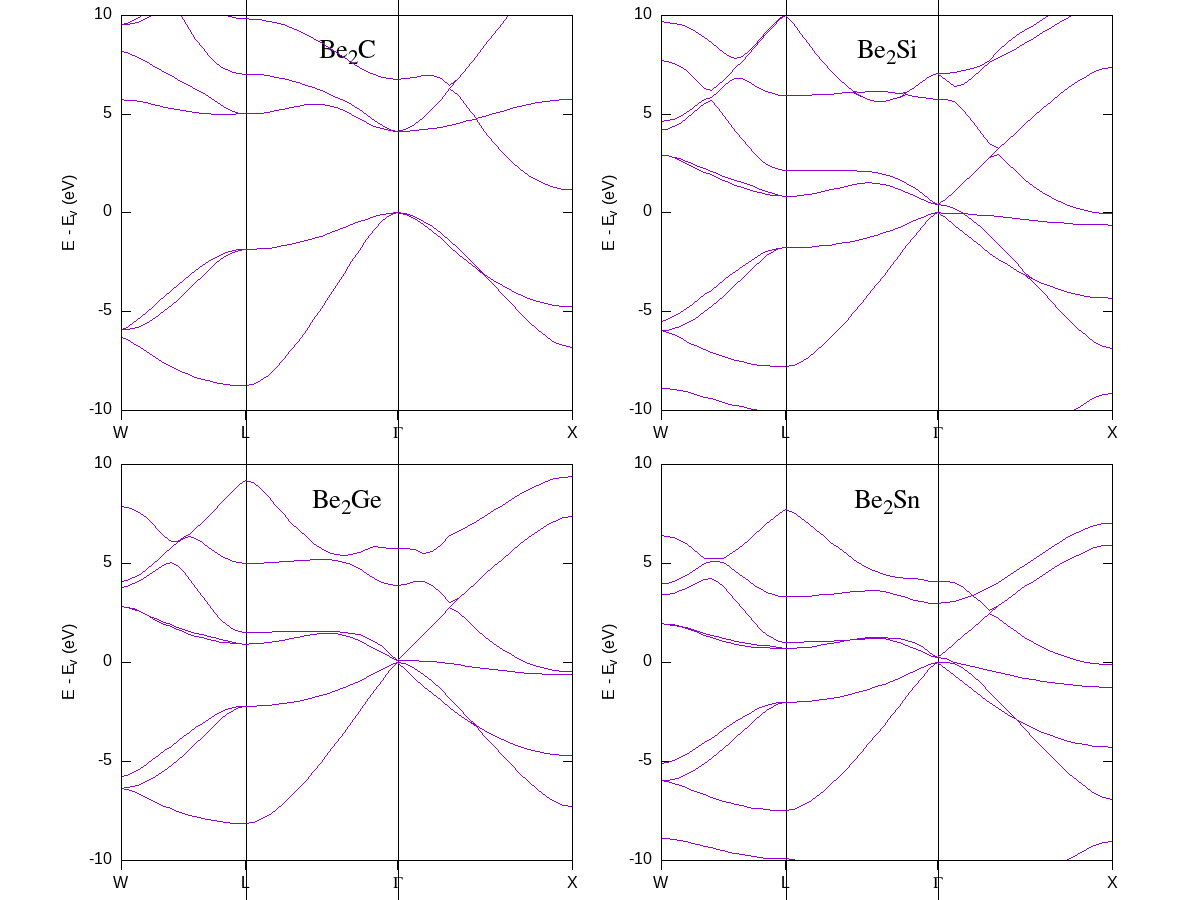}
	\caption{Band Structure of $Be_2X(X=C, Si, Ge, Sn)$ Crystal. Valence Band Maximum is set to 0 eV}
\end{figure}
The bandgap was also calculated using different hybrid functional with 8x8x8 k-grid and 8x8x8 q-grid shown in Table 5.
\begin{table}[h]
	\centering
	\caption{Bandgap values using different hybrid functionals}
	\begin{tabular}{p{3cm}p{2cm}p{2cm}p{2cm}}
		\toprule
		& GGA [eV] & PBE0 [eV] & HSE [eV]  \\
		\midrule
		$Be_2C$ & 1.1701 & 2.5569  & 1.8885 \\
		$Be_2Si$ & 0.8142* & 0.1067*  & 0.5766*  \\
		$Be_2Ge$ & 0.6218* & 0.601  & 0.0352  \\
		$Be_2Sn$  & 1.3015* & 0.8821*  & 1.2888*   \\
		\bottomrule
	\end{tabular} \\
	\raggedright
	* Band Crossing
\end{table}
\begin{figure}[!h]
	\centering
	\includegraphics[width=1.1\linewidth]{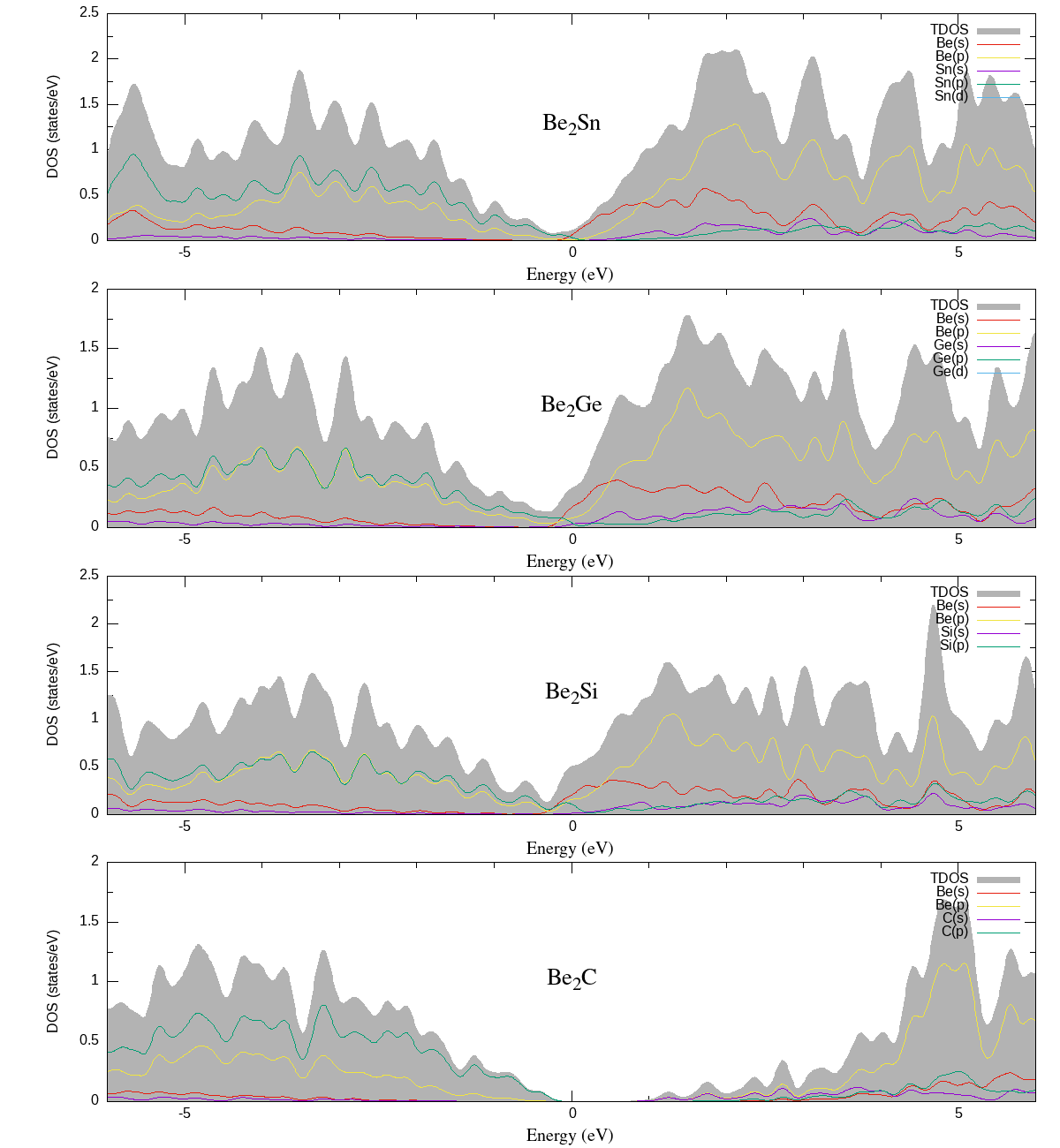}
	\caption{Density of states of $Be_2X(X=C, Si, Ge, Sn)$ Crystal. Valence Band Maximum is set to 0 eV}
\end{figure} \\
\newpage
\subsection*{Optical Properties}
The dielectric function give the behaviour of the material that how it responds to the incident electromagnetic radiation. We calculated the real part $\varepsilon_1(\omega)$ and imaginary part $\varepsilon_2(\omega)$ of the dielectric function by solving Bethe-Salpeter Equations. The real part that shows the physical properties of a crystal. The imaginary part shows the energy loss of photons in a material when there is electron transition between electronic bands. The real and imaginary are plotted in Figure 4.
\begin{figure}[h!]
	\includegraphics[width=1.1\linewidth]{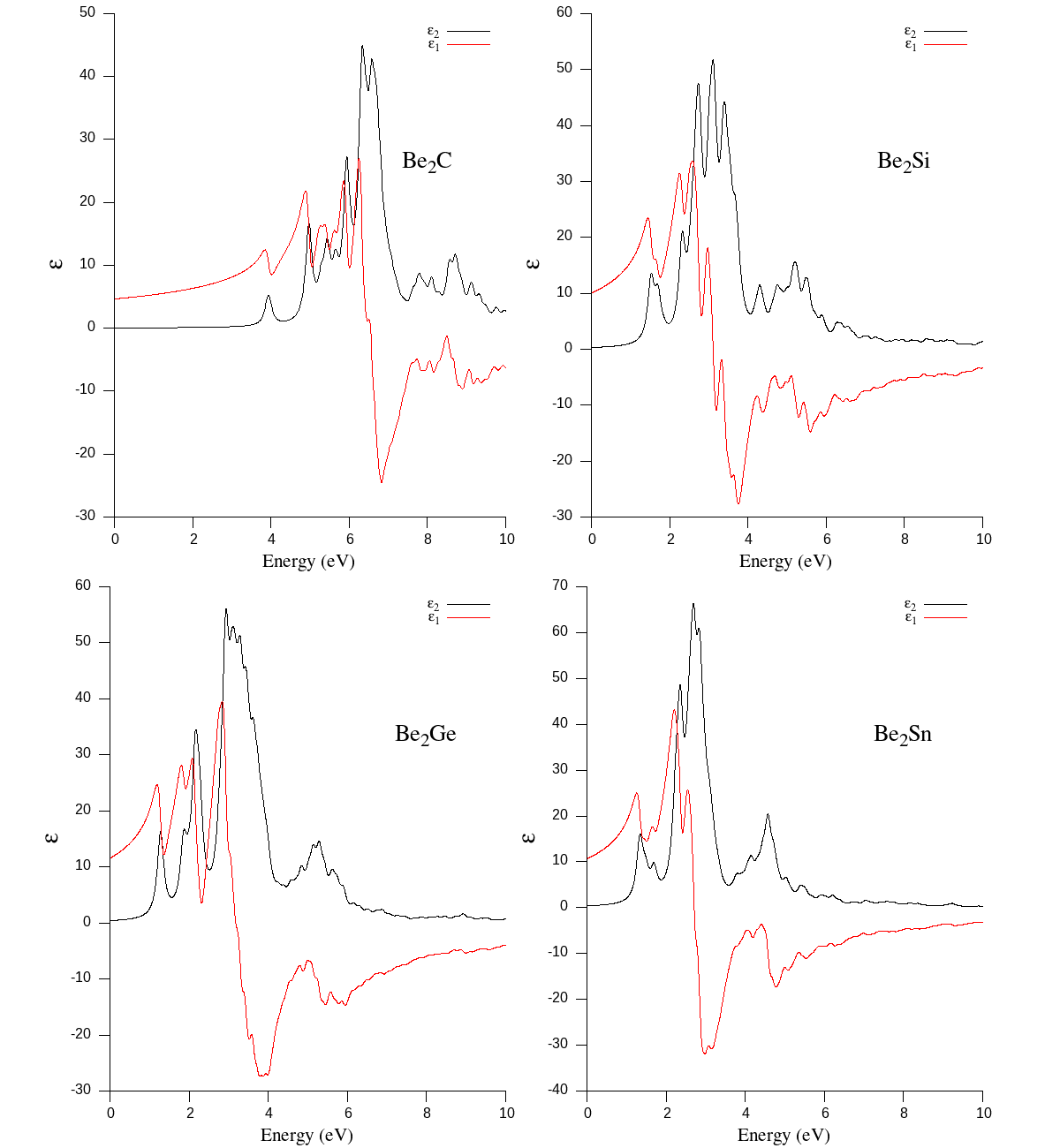}
	\caption{Real part $\varepsilon_1(\omega)$ and imaginary part $\varepsilon_2(\omega)$ of the dielectric function.}
\end{figure} \\
In real part of dielectric function graphs, the regions where real dielectric function decrease with the increase of photon energy gives the abnormal dispersion region. The points where $\varepsilon$ value takes zero value at some phonon energies corresponds to plasmon frequencies \cite{MOCHAN2005310}. Similarly, the regions where $\varepsilon$ values become negative indicates that there is a reflection of the incident electromagnetic radiations. The plasmon frequencies and the negative regions of real part of $\varepsilon$ can be obeserved from Figure 4. The calculated value of static dielectric constant, $\varepsilon_1(0)$,  for $Be_2X$ is shown in Table 6. In the imaginary part of dielectric  function graphs, the peaks corresponds to the transitions of electron from valence to conduction bands \cite{erdinc2019investigation}. \\
\begin{table}[h]
	\centering
	\caption{Static dielectric constants $\varepsilon_1(0)$ for $Be_2X(X=C, Si, Ge, Sn)$}
	\begin{tabular}{p{3cm}p{3cm}p{2cm}}
		\toprule
		& $\varepsilon_1(0) ^p$  & $\varepsilon_1(0) ^q$ \\
		\midrule
		$Be_2C$ & 4.65 & 4.11  \\
		$Be_2Si$ & 10.04  & 9.75  \\
		$Be_2Ge$ & 11.53  & 10.70  \\
		$Be_2Sn$ & 10.68  & 10.21  \\
		\bottomrule
	\end{tabular} \\
	\raggedright
	\vspace{0.5cm}
	$^p$ solving BSE \\
	$^q$ with IPA \\
\end{table} \\
Some other optical property terms that includes refractive index $n(\omega)$, extinction coefficient $k(\omega)$, absorption coefficient $\alpha (\omega)$, reflectivity $R(\omega)$ and energy loss spectra $L(\omega)$ are calculated as \cite{doi:10.1063/1.3499308}: \\
\begin{align*}
	n(\omega) &= \frac{1}{\sqrt{2}} \left[ \sqrt{\varepsilon_1 ^2 (\omega) + \varepsilon_2^2(\omega)} + \varepsilon_1(\omega) \right] ^{1/2} \\ \\
	k(\omega) &= \frac{1}{\sqrt{2}} \left[ \sqrt{\varepsilon_1 ^2 (\omega) + \varepsilon_2^2(\omega)} - \varepsilon_1(\omega) \right] ^{1/2} \\ \\
	\alpha (\omega) &= \sqrt{2} \frac{\omega}{c} \left[ \sqrt{\varepsilon_1 ^2 (\omega) + \varepsilon_2^2(\omega)} - \varepsilon_1(\omega) \right] ^{1/2} \\ \\
	R(\omega) &= \frac{(n(\omega) - 1)^2 + k^2(\omega)}{(n(\omega) + 1)^2 + k^2(\omega)} \\ \\
	L(\omega) &= \frac{\varepsilon_2(\omega)}{\varepsilon_1 ^2 (\omega) + \varepsilon_2^2(\omega)} \\
\end{align*}
They are shown in Appendices. The refractive index of the four structures is shown in Table 6 which satisfies the relation $n(0) = \sqrt{\varepsilon_1(0)}$. The larger value of refractive index of the materials indicates that when photons enters the material, they are slowed down due to interaction with electrons. \\
\begin{table}[h]
	\centering
	\caption{Calculated Optical parameters of $Be_2X(X=C,Si,Ge,Sn)$}
	\begin{tabular}{p{4cm}p{1.5cm}p{1.5cm}p{1.5cm}p{1.5cm}}
		\toprule
		& $Be_2C$ & $Be_2Si$ & $Be_2Ge$ & $Be_2Sn$ \\
		\midrule
		\midrule
		$n(0)$ &  2.15 & 3.169 & 3.39 & 3.26 \\
		$\varepsilon_1(0)$ & 4.65 & 10.04 & 11.53 & 10.68 \\
		Maximum $n(\omega)$ & 5.93 & 6.37 & 6.97 & 6.91 \\
		$R(0)$ \% & 13.45 & 27.08 & 29.73 & 28.27 \\
		Maximum $R(\omega)$ \% & 82.09 & 86.31 & 92.78 & 94.19  \\
		\bottomrule
	\end{tabular}
\end{table}\\
The zero-frequency reflectivity $R(0)$ and maximum reflectivity $n(\omega)$ for the four structures are mentioned in the Table 7. The reflectivity shows high reflectivity in the UV region. The absorbance spectra shows low absorbance in Visible region and high absorbance is found in the UV region for all four structures. \\
\newpage
\section{Conclusion}
A first principle DFT method has been implemented to investigate Structure, Elastic, Electronic and Optical properties of $Be_2X(X=C, Si, Ge, Sn)$. The lattice parameter at equilibrium is in good agreement with experimental data and previously calculated for $Be_2C$ and $Be_2Si$. The study of elastic property indicates fragile and brittle nature of $Be_2C$. The band gap in cubic structure of $Be_2C$ is found to be 4.132 eV. The other three structures $Be_2X(X=Si, Ge, Sn)$ shows metallicity nature. The optical properties terms that includes dielectric function, reflectivity, absorption coeffcient, refractive index, extinction coeffcient, and electron energy loss are studied in 0-10 eV energy range. The negative value of $\varepsilon$ shows that $Be_2C$ and other three structures can be used as blanket material for prevention from UV radiation damage. The low optical absorbance of $Be_2C$ against other three structures in the visible region restricts its use in Solar Cell and other optopelectronic applications.  Due to less research work on $Be_2C$ and hypothetical compounds $Be_2X(X=Si, Ge, Sn)$, all results has not be compared. Thus, this study can be helpful for further research on this crystal.
\section{Acknowledgement}
This work was supported in part through computational resources
provided by the Supercomputer Center Kathmandu University, which was
established with equipments donated by CERN. \\ 
\bibliography{Be2X}

\begin{thebibliography}{10}
\providecommand{\url}[1]{\normalfont{#1}}
\providecommand{\urlprefix}{Available from: }

\bibitem{C39890001652}
Fowler~PW, Tole~P. {Theoretical evidence for the C4- ion in beryllium carbide}.
  Journal of the Chemical Society, Chemical Communications.
  1989;\hspace{0pt}(21):1652--1654.
  \urlprefix\url{http://dx.doi.org/10.1039/C39890001652}.

\bibitem{C9RA01573F}
Maurya~V, Paliwal~U, Sharma~G, et~al. {Thermoelectric and vibrational
  properties of Be2C, BeMgC and Mg2C using first-principles method}. RSC Adv.
  2019;\hspace{0pt}9(24):13515--13526.
  \urlprefix\url{http://dx.doi.org/10.1039/C9RA01573F}.

\bibitem{1954}
Mallett~MW, Durbin~EA, Udy~MC, et~al. {Preparation and Examination of Beryllium
  Carbide}. Journal of The Electrochemical Society.
  1954;\hspace{0pt}101(6):298.
  \urlprefix\url{https://doi.org/10.1149/1.2781251}.

\bibitem{1952}
Coobs~JH, Koshuba~WJ. {The Synthesis, Fabrication, and Properties of Beryllium
  Carbide}. Journal of The Electrochemical Society. 1952;\hspace{0pt}99(3):115.
  \urlprefix\url{https://doi.org/10.1149/1.2779672}.

\bibitem{KLEYKAMP200188}
Kleykamp~H. {Selected thermal properties of beryllium and phase equilibria in
  beryllium systems relevant for nuclear fusion reactor blankets}. Journal of
  Nuclear Materials. 2001;\hspace{0pt}294(1-2):88--93.
  \urlprefix\url{https://www.sciencedirect.com/science/article/pii/S0022311501004421}.

\bibitem{771086}
Hite~D, Tang~SJ, Sprunger~P. {Reactive epitaxy of beryllium on Si(1 1 1)-(7 x
  7)}. Chemical Physics Letters. 2003;\hspace{0pt}(367).

\bibitem{article}
Saranin~A, Zotov~A, Kotlyar~VG, et~al. {Self-assembly formation of the ordered
  nanostructure arrays induced by Be interaction with Si(111) surface}. Surface
  Science. 2005;\hspace{0pt}574:99--109.

\bibitem{sholl2011density}
Sholl~D, Steckel~JA. {Density functional theory: a practical introduction}.
  John Wiley {\&} Sons; 2011.

\bibitem{singh2006planewaves}
Singh~DJ, Nordstrom~L. {Planewaves, Pseudopotentials, and the LAPW method}.
  Springer Science {\&} Business Media; 2006.

\bibitem{PhysRevLett.77.3865}
Perdew~JP, Burke~K, Ernzerhof~M. {Generalized Gradient Approximation Made
  Simple}. Phys Rev Lett. 1996 oct;\hspace{0pt}77(18):3865--3868.
  \urlprefix\url{https://link.aps.org/doi/10.1103/PhysRevLett.77.3865}.

\bibitem{doi:10.1063/1.472933}
Perdew~JP, Ernzerhof~M, Burke~K. {Rationale for mixing exact exchange with
  density functional approximations}. The Journal of Chemical Physics.
  1996;\hspace{0pt}105(22):9982--9985.
  \urlprefix\url{https://doi.org/10.1063/1.472933}.

\bibitem{doi:10.1063/1.1564060}
Heyd~J, Scuseria~GE, Ernzerhof~M. {Hybrid functionals based on a screened
  Coulomb potential}. The Journal of Chemical Physics.
  2003;\hspace{0pt}118(18):8207--8215.
  \urlprefix\url{https://doi.org/10.1063/1.1564060}.

\bibitem{GOLESORKHTABAR20131861}
Golesorkhtabar~R, Pavone~P, Spitaler~J, et~al. {ElaStic: A tool for calculating
  second-order elastic constants from first principles}. Computer Physics
  Communications. 2013;\hspace{0pt}184(8):1861--1873.
  \urlprefix\url{https://www.sciencedirect.com/science/article/pii/S0010465513001070}.

\bibitem{PhysRevB.48.11705}
Sipe~JE, Ghahramani~E. {Nonlinear optical response of semiconductors in the
  independent-particle approximation}. Phys Rev B. 1993
  oct;\hspace{0pt}48(16):11705--11722.
  \urlprefix\url{https://link.aps.org/doi/10.1103/PhysRevB.48.11705}.

\bibitem{2019}
Sangalli~D, Ferretti~A, Miranda~H, et~al. {Many-body perturbation theory
  calculations using the yambo code}. 2019 may;\hspace{0pt}31(32):325902.
  \urlprefix\url{https://doi.org/10.1088/1361-648x/ab15d0}.

\bibitem{RUSCHEWITZ2003115}
Ruschewitz~U. {Binary and ternary carbides of alkali and alkaline-earth
  metals}. Coordination Chemistry Reviews. 2003;\hspace{0pt}244(1-2):115--136.
  \urlprefix\url{https://www.sciencedirect.com/science/article/pii/S0010854503001024}.

\bibitem{Murnaghan1944}
Murnaghan~FD. {The Compressibility of Media under Extreme Pressures.}
  Proceedings of the National Academy of Sciences of the United States of
  America. 1944 sep;\hspace{0pt}30(9):244--247.

\bibitem{doi:10.1063/1.4810267}
Paliwal~U, Trivedi~DK, Galav~KL, et~al. {First-principles study of structural
  properties of alkaline earth metals methanides A2C(A = Be,Mg)}. AIP
  Conference Proceedings. 2013;\hspace{0pt}1536(1):395--396.
  \urlprefix\url{https://aip.scitation.org/doi/abs/10.1063/1.4810267}.

\bibitem{Yan2011}
Yan~HY, Wei~Q, Chang~SM, et~al. {A First Principle Study of Antifluorite Be 2 X
  (X = C, Si) Polymorph}. Acta Physica Polonica A. 2011
  mar;\hspace{0pt}119:442--446.

\bibitem{PhysRevB.51.10392}
Lee~CH, Lambrecht~WR, Segall~B. {Electronic structure of Be2C}. Physical Review
  B. 1995 apr;\hspace{0pt}51(16):10392--10398.
  \urlprefix\url{https://link.aps.org/doi/10.1103/PhysRevB.51.10392}.

\bibitem{PhysRevB.48.17138}
Corkill~JL, Cohen~ML. {Structural, bonding, and electronic properties of IIA-IV
  antifluorite compounds}. Phys Rev B. 1993
  dec;\hspace{0pt}48(23):17138--17144.
  \urlprefix\url{https://link.aps.org/doi/10.1103/PhysRevB.48.17138}.

\bibitem{PhysRev.136.B864}
Hohenberg~P, Kohn~W. {Inhomogeneous Electron Gas}. Phys Rev. 1964
  nov;\hspace{0pt}136(3B):B864----B871.
  \urlprefix\url{https://link.aps.org/doi/10.1103/PhysRev.136.B864}.

\bibitem{Man2011}
Man~CS, Huang~M. {A Simple Explicit Formula for the Voigt-Reuss-Hill Average of
  Elastic Polycrystals with Arbitrary Crystal and Texture Symmetries}. Journal
  of Elasticity. 2011;\hspace{0pt}105(1):29--48.
  \urlprefix\url{https://doi.org/10.1007/s10659-011-9312-y}.

\bibitem{Varshney2015}
Varshney~D, Shriya~S, Varshney~M, et~al. {Elastic and thermodynamical
  properties of cubic (3C) silicon carbide under high pressure and high
  temperature}. Journal of Theoretical and Applied Physics.
  2015;\hspace{0pt}9(3):221--249.
  \urlprefix\url{https://doi.org/10.1007/s40094-015-0183-7}.

\bibitem{PhysRevB.90.224104}
Mouhat~F, Coudert~FX. {Necessary and sufficient elastic stability conditions in
  various crystal systems}. Phys Rev B. 2014 dec;\hspace{0pt}90(22):224104.
  \urlprefix\url{https://link.aps.org/doi/10.1103/PhysRevB.90.224104}.

\bibitem{harrison2012electronic}
Harrison~WA. {Electronic structure and the properties of solids: the physics of
  the chemical bond}. Courier Corporation; 2012.

\bibitem{PhysRevB.58.6837}
Tzeng~CT, Tsuei~KD, Lo~WS. {Experimental electronic structure of Be2C}. Phys
  Rev B. 1998 sep;\hspace{0pt}58(11):6837--6843.
  \urlprefix\url{https://link.aps.org/doi/10.1103/PhysRevB.58.6837}.

\bibitem{MOCHAN2005310}
Mochan~WL. {Plasmons}. In: Bassani~F, Liedl~GL, Wyder~P, editors. Encyclopedia
  of condensed matter physics. Oxford: Elsevier; 2005. p. 310--317.
  \urlprefix\url{https://www.sciencedirect.com/science/article/pii/B0123694019006616}.

\bibitem{erdinc2019investigation}
Erdinc~F, Dogan~EK, Akkus~H. {Investigation of Structural, Electronic, Optic
  and Elastic Properties of Perovskite RbGeCl3 Crystal: A First Principles
  Study}. Gazi University Journal of Science.
  2019;\hspace{0pt}32(3):1008--1019.

\bibitem{doi:10.1063/1.3499308}
Sun~L, Zhao~X, Li~Y, et~al. {First-principles studies of electronic, optical,
  and vibrational properties of LaVO4 polymorph}. Journal of Applied Physics.
  2010;\hspace{0pt}108(9):93519.
  \urlprefix\url{https://doi.org/10.1063/1.3499308}.

\end{thebibliography}
\newpage
\appendix
\section{ Total Energy with respect to volume (au$^3$)}
\begin{figure}[h!]
	\centering
	\includegraphics[width=1\linewidth]{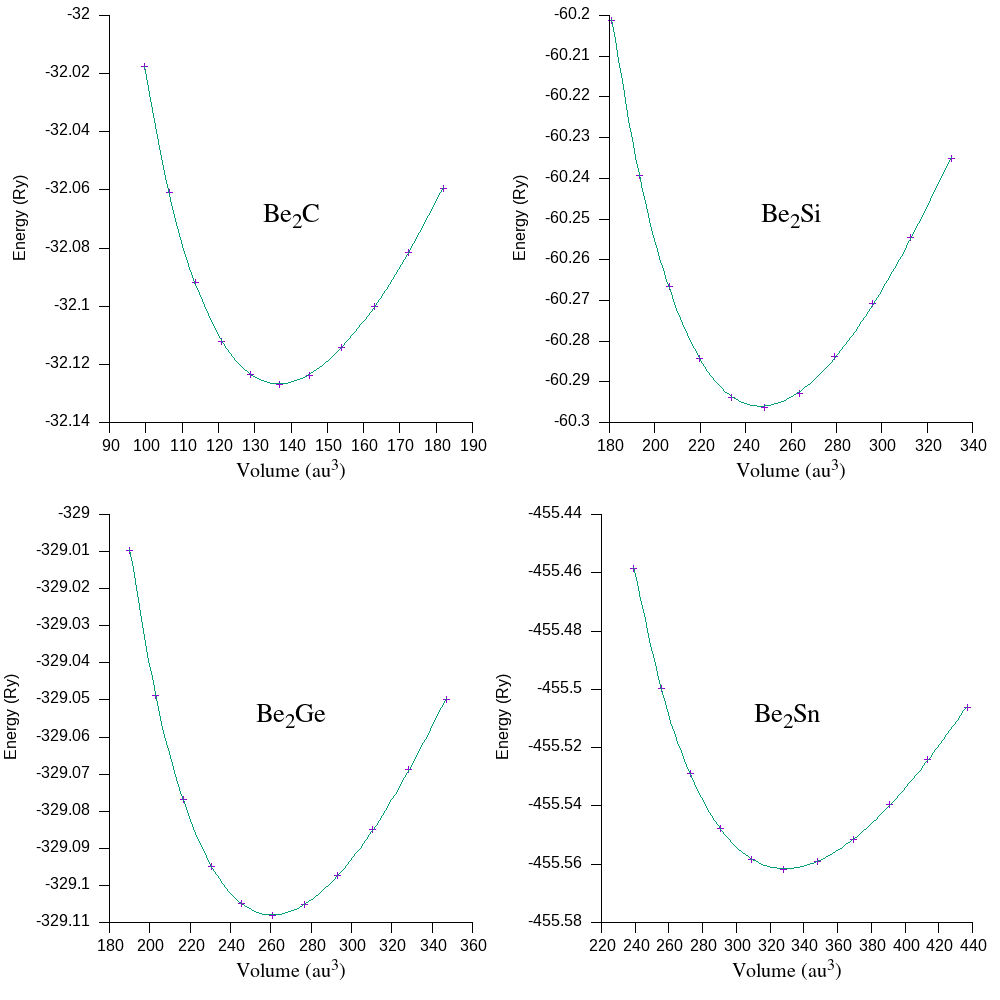}
\end{figure}
\newpage
\section{Calculated refractive index, extinction coefficient, absorption coefficient, reflectivity and Energy loss spectra of $Be_2C$.}
\begin{figure}[!ht]
	\centering
	\includegraphics[width=0.95\linewidth]{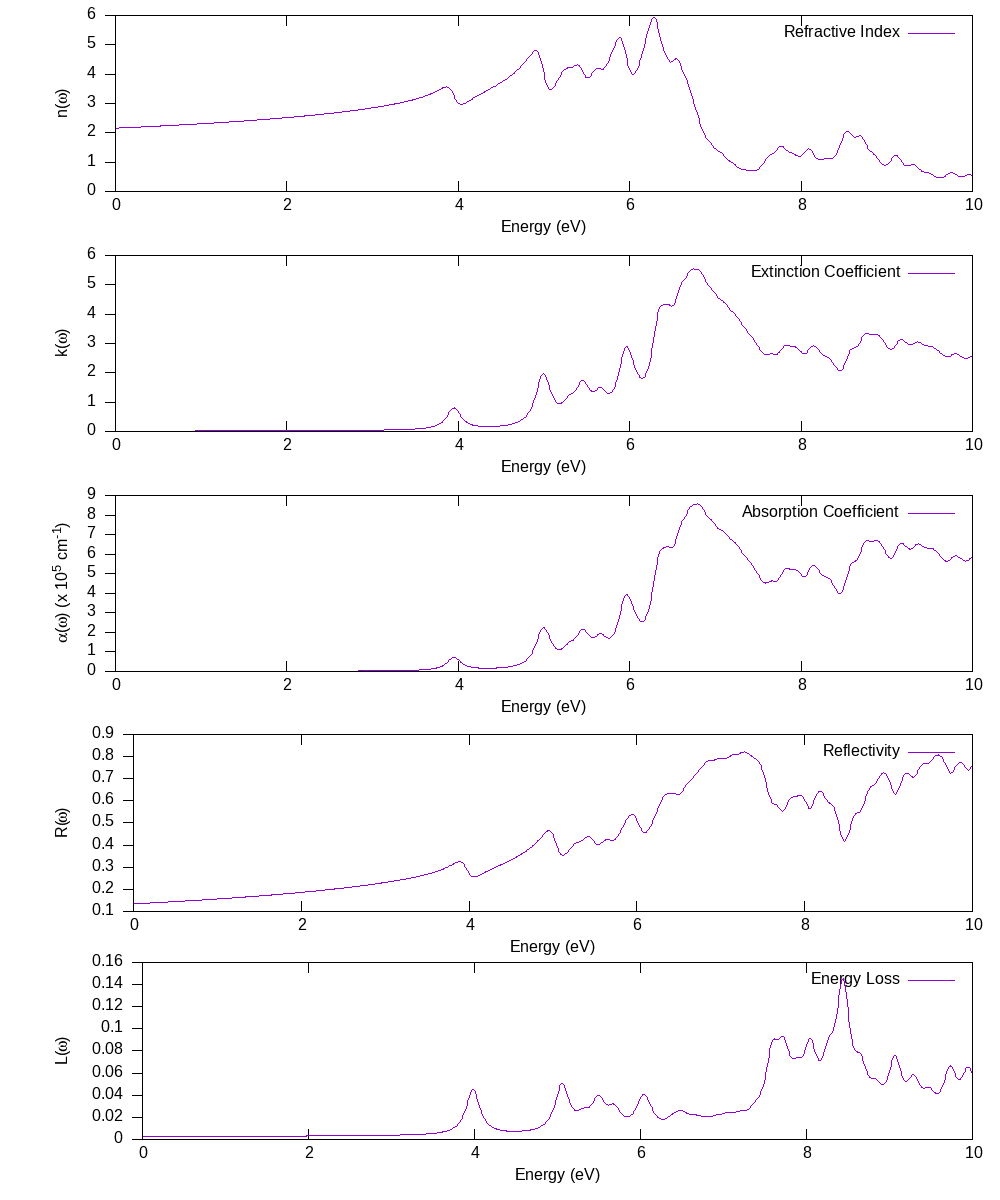}
\end{figure}
\newpage
\section{Calculated refractive index, extinction coefficient, absorption coefficient, reflectivity and Energy loss spectra of $Be_2Si$.}
\begin{figure}[!ht]
	\centering
	\includegraphics[width=0.95\linewidth]{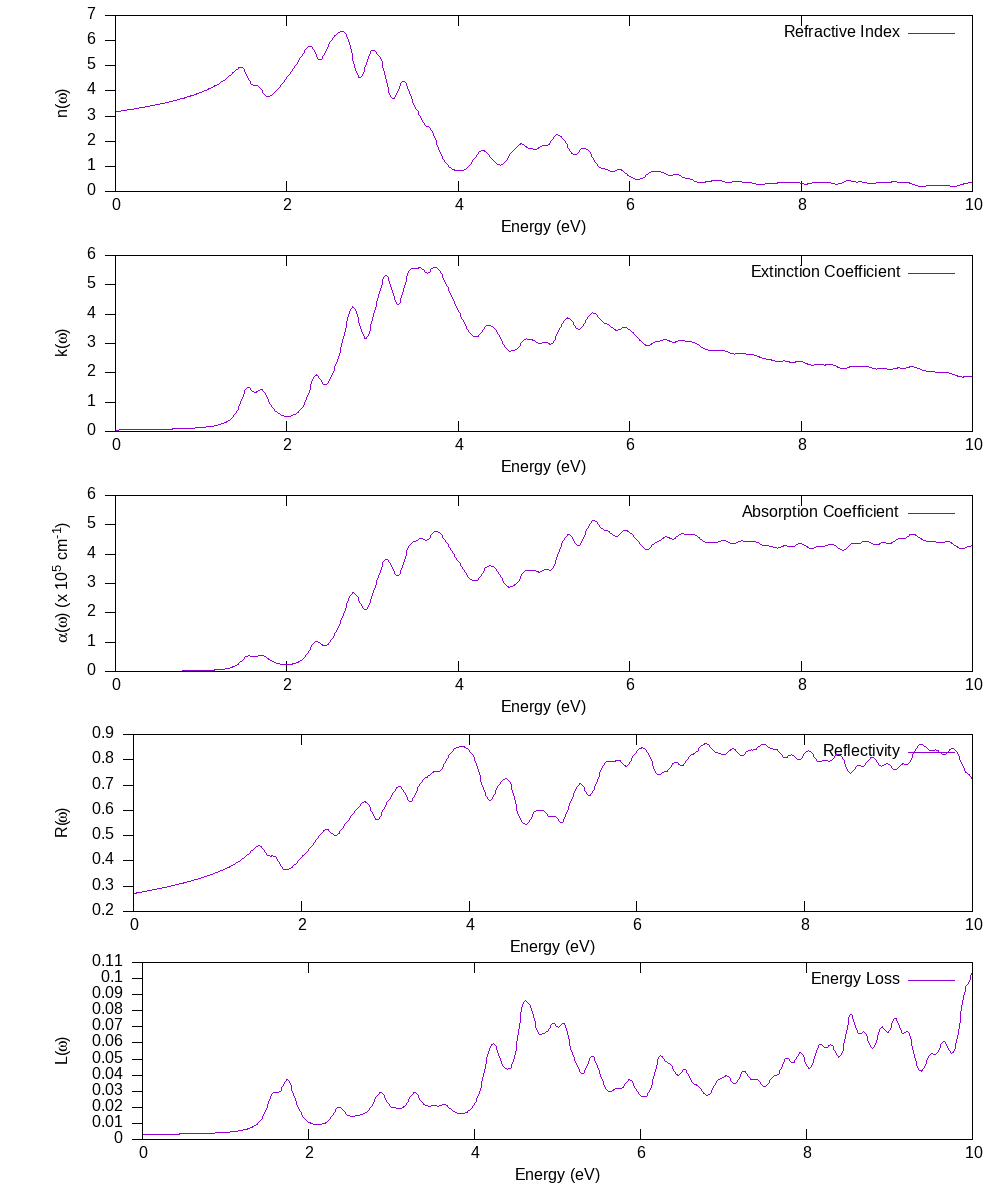}
\end{figure}
\newpage
\section{Calculated refractive index, extinction coefficient, absorption coefficient, reflectivity and Energy loss spectra of $Be_2Ge$.}
\begin{figure}[!ht]
	\centering
	\includegraphics[width=0.95\linewidth]{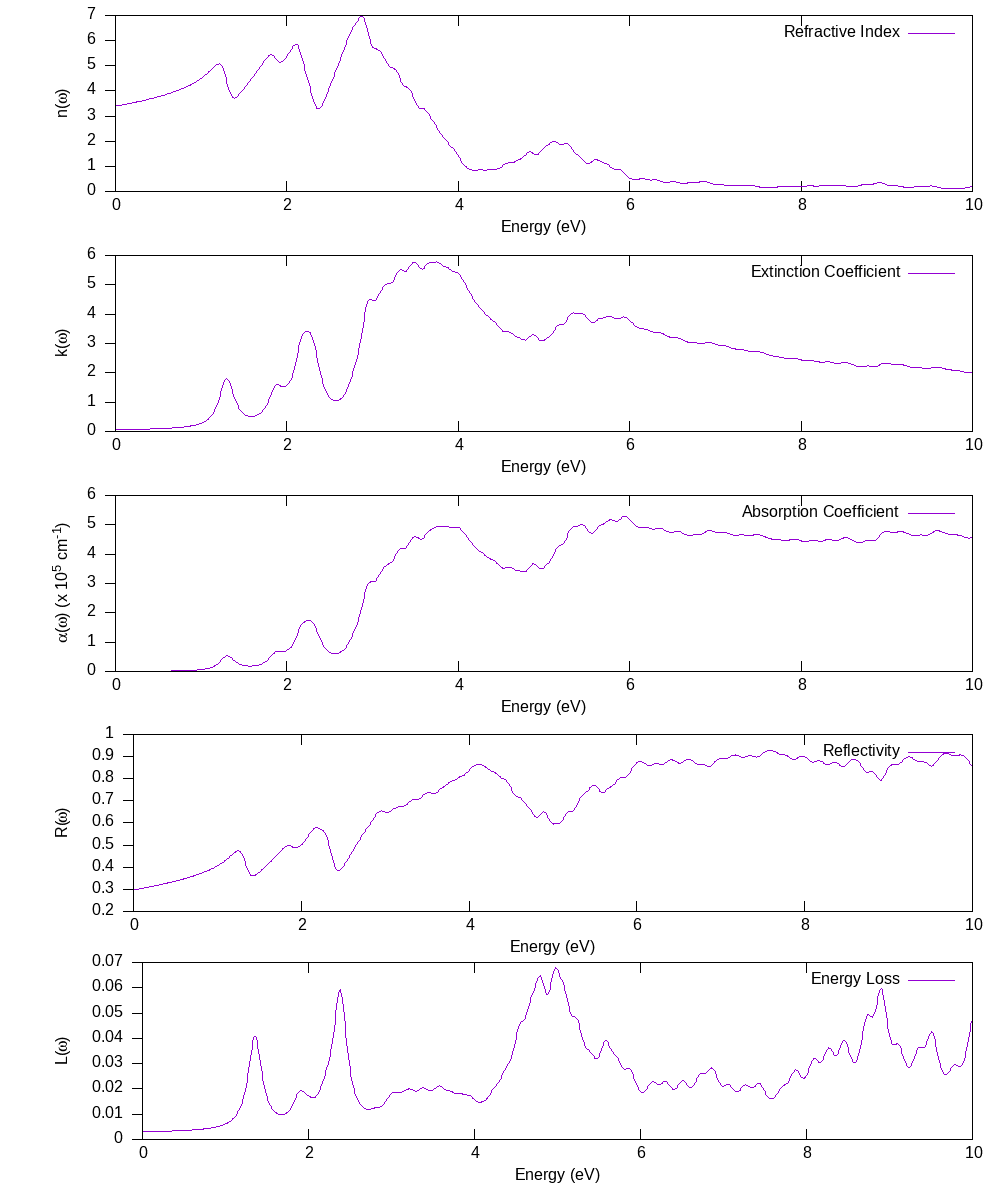}
\end{figure}
\newpage
\section{Calculated refractive index, extinction coefficient, absorption coefficient, reflectivity and Energy loss spectra of $Be_2Sn$.}
\begin{figure}[!ht]
	\centering
	\includegraphics[width=0.95\linewidth]{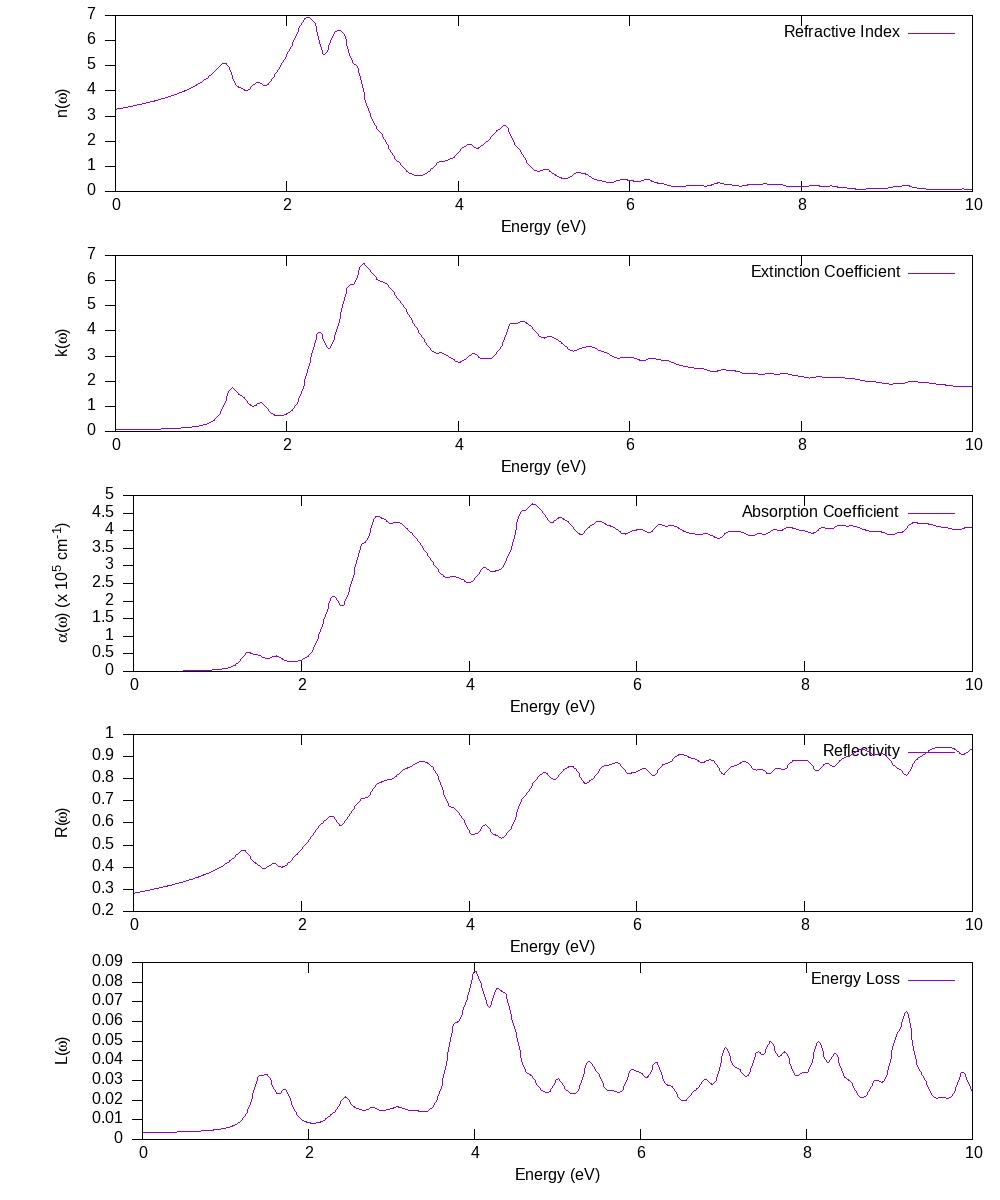}
\end{figure}

\end{document}